\documentclass{JINST}

\usepackage{amssymb}
\usepackage{amsmath}
\usepackage{xspace}
\usepackage{textcomp}
\usepackage{cite}

\title{Multiple Coulomb Scattering in Thin Silicon}

\author{Niklaus Berger$^a$\thanks{Corresponding
author.}, Armen Buniatyan$^a$, Patrick Eckert$^b$, Fabian F\"orster$^a$, Roman Gredig$^c$, Oxana Kovalenko$^{a,d,e}$, Moritz Kiehn$^a$, Raphael Philipp$^a$, Andr\'e Sch\"oning$^a$  and Dirk Wiedner$^a$\\
\llap{$^a$}Physikalisches Institut,\\
  Heidelberg University, Heidelberg, Germany\\
\llap{$^b$}Kirchhoff-Institut f\"ur Physik,\\
  Heidelberg University, Heidelberg, Germany\\
\llap{$^c$}Physik Institut,\\
  Z\"urich University, Z\"urich, Switzerland\\
\llap{$^d$}Budker Institute of Nuclear Physics,\\
  Novosibirsk, Russia\\	
\llap{$^e$}Novosibirsk State University,\\
  Novosibirsk, Russia\\		
  E-mail: \email{nberger@physi.uni-heidelberg.de}}

\abstract{We present a measurement of multiple Coulomb scattering of 1 to 6~GeV/$c$ electrons in thin (50-140~$\mu$m) silicon targets. The data were obtained with the EUDET telescope \emph{Aconite} at DESY and are compared to parametrisations as used in the Geant4 software package. We find good agreement between data and simulation in the scattering distribution width but large deviations in the shape of the distribution. In order to achieve a better description of the shape, a new scattering model based on a Student's $t$ distribution is developed and compared to the data.}


\begin{document}

\section{Motivation}

A good understanding of multiple Coulomb scattering of relativistic particles in matter is important both for tracking
detectors and calorimetry. The theory of multiple scattering was first treated by Wentzel in 1922 \cite{Wentzel1922}
and fully developed in the 1940ies by Goudsmit and Saunderson
\cite{Goudsmit:1940zza, Goudsmit:1940zz} and Moli\`ere \cite{Moliere:1947, Moliere:1948zz} (summarized in more
elegant notation by Bethe \cite{Bethe:1953va}). Their approaches differ in the treatment of the screened nuclear
potential and the series expansion applied to make the problem analytically tractable. In both calculations however, the path 
length of the particle in the material is assumed to be independent of the scattering angle; this problem was addressed
by Lewis \cite{Lewis:1950zz}, whose improved approach is also the basis of the default multiple scattering model in the Geant4
simulation package \cite{Allison:2012gn, Ivanchenko:2010zz}. For an extensive review of multiple scattering theory, see
\cite{Scott1963}. 

For experimental purposes, very often the parametrisation suggested by Highland \cite{Highland:1975pq}
and popularized by the Particle Data Group (PDG) \cite{PDG2012} is used:
\begin{equation}
	\theta_0 = \frac{13.6~\mathrm{MeV}}{\beta c p} \; z \;\sqrt{\frac{x}{X_0}}\left( 1 + 0.038 \ln \left(\frac{x}{X_0}\right) \right),
\end{equation}
where $\theta_0$ is the $RMS$ calculated from the central 98\% of the planar scattering angle distribution (henceforth referred to as $RMS_{98}$), $p$, $\beta c$ and $z$ are the momentum, velocity and charge number of the incident particle and $x/X_0$ is the material thickness in radiation lengths.

Whilst there is a wealth of data on multiple scattering measured in silicon of a few 100~$\mu$m thickness in solid state 
tracking devices, there are only few
published data for scattering in thin foils. Measurements of the scattering of 15.7~MeV/$c$ electrons in beryllium 
by Hanson \emph{et al.} \cite{Hanson:1951zz} disagreed with the theoretical models available at the time, whilst the measurements 
with a gold foil were adequately described. They serve as 
benchmarks for models to this day \cite{Ivanchenko:2010zz}. Other published datasets range from $2.25$~MeV/$c$ electrons \cite{Kulchitsky:1942zz} 
via measurements with pions at a few hundred MeV/$c$ \cite{Mayes:1974gs} to protons with 0.7 and 4.8~MeV/$c$ \cite{Bichsel1958}, 600~MeV~\cite{Hungerford:1973ky}
and 50 to 200~GeV/$c$ \cite{Shen:1978ha}. As targets, metal and carbon foils and slabs were used.

With the advent of very thin (50~$\mu$m) silicon tracking detectors \cite{Kemmer:1986vh,Turchetta:2001dy,Neese:2002zz,Winter2010192,Ulrici:2000hv,DEPFET:2007zzm,Peric:2007zz, Peric:2013cka},
a good understanding of scattering in thin silicon is important for the design and calibration of experiments employing such sensors, such as the STAR
pixel detector \cite{Dorokhov:2011zzb, Margetis:2011zz}, the BELLE II silicon tracker \cite{ DEPFET:2007zzm,Abe:2010sj, Kreidl2012} or the Mu3e pixel detector \cite{RP, Berger:2013raa}. The present measurement is performed in the course of a test beam campaign for the Mu3e experiment.   

\section{Measurement Setup}

\begin{figure*}
\centering 
\includegraphics[width=0.9\textwidth]{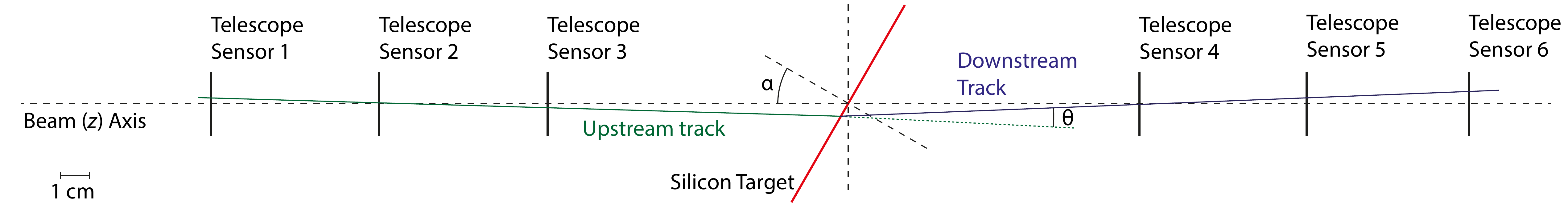}
\caption{Scale drawing of the setup for the multiple scattering measurement with the \emph{Aconite} EUDET telescope, top view. The electron beam enters from the left. The incidence angle of the beam is denoted by $\alpha$ and the scattering angle of the electron by $\theta$.}
\label{fig:setup}
\end{figure*}

The data presented here were obtained with the EUDET telescope \emph{Aconite} \cite{Gregor:2008zz, Roloff:2009zza, Haas:2006vw} at the DESY test beam line T22. The beam line provides electrons from converted bremsstrahlung beams produced by carbon fibre targets in the electron synchrotron DESY II with momenta from 1 to 6~GeV/$c$ and an energy spread below 5\% \cite{Autiero2004} at rates up to about 1~kHz. The beam divergence is approximately 1~mrad.

The telescope is built from six layers of Mimosa26 \cite{Dorokhov:2011zzb, Baudot:2009dta, Valin:2012zz} monolithic active pixel sensors (MAPS) thinned to 50~$\mu$m. The active area of the MAPS is approximately $2 \times 1$~cm$^2$. The data acquisition is triggered by a coincidence of signals in two crossed pairs of scintillators, one before and one after the telescope. Between the third and fourth telescope plane, we placed either one or two 50~$\mu$m thick unprocessed silicon wafers\footnote{The manufacturer specifies the wafer thickness as 40-60~$\mu$m; we measured 50~$\mu$m within an uncertainty of 5~$\mu$m for all samples.} as scattering targets on a rotating stage, see Figure~\ref{fig:setup} for an overview of the set-up. The silicon wafers are much larger than the Mimosa sensors, it is thus ensured that all tracks in the telescope acceptance pass through the target. The sixth telescope plane was out of operation for this measurement.

Data were taken at electron momenta between 1 and 6~GeV/$c$ in 1~GeV/$c$ increments. 
For every momentum point, we measured scattering angles with a 50 or 100~$\mu$m thick target oriented at beam incidence angles of 0$^\circ$, 15$^\circ$, 30$^\circ$ and 45$^\circ$ resulting in a projected thickness $d_{\textnormal{eff}}$ between 50 and 141~$\mu$m. For a determination of the contribution of the telescope, measurements without the silicon target were performed.
For each data point we collected about one million triggers, resulting in approximately 300'000 tracks after selection cuts.

\section{Data Analysis}

\begin{figure}[b!]
	\centering
		\includegraphics[width=0.46\textwidth]{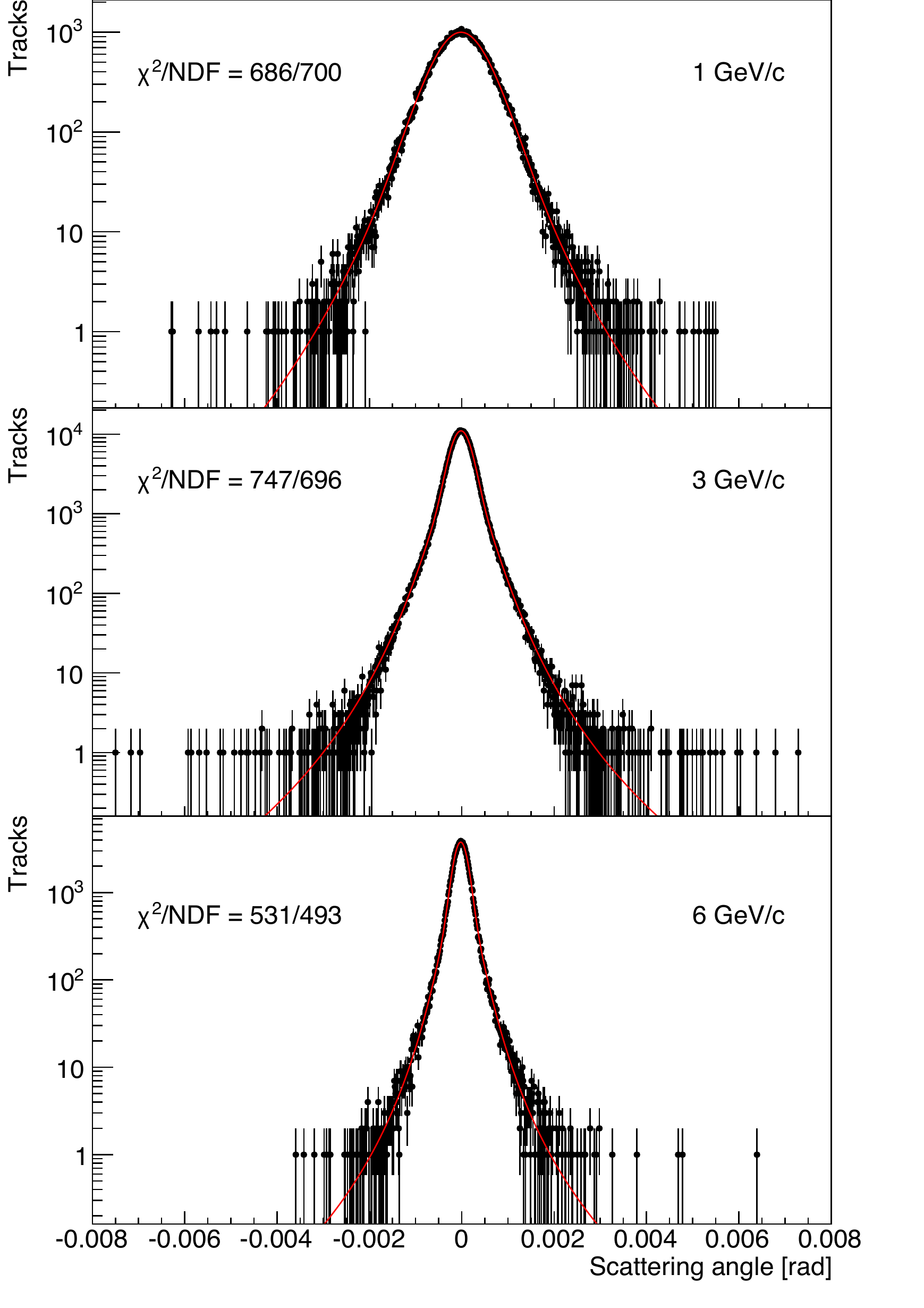}
	\caption{Horizontal scattering angle distribution for 1, 3 and 6~GeV/c electrons with no scattering target. As fit function (red line), the sum of a Gaussian and a Student's $t$-distribution is used as described in the text.}
	\label{fig:nosilicon}
\end{figure}

\begin{figure}[b!]
	\centering
		\includegraphics[width=0.46\textwidth]{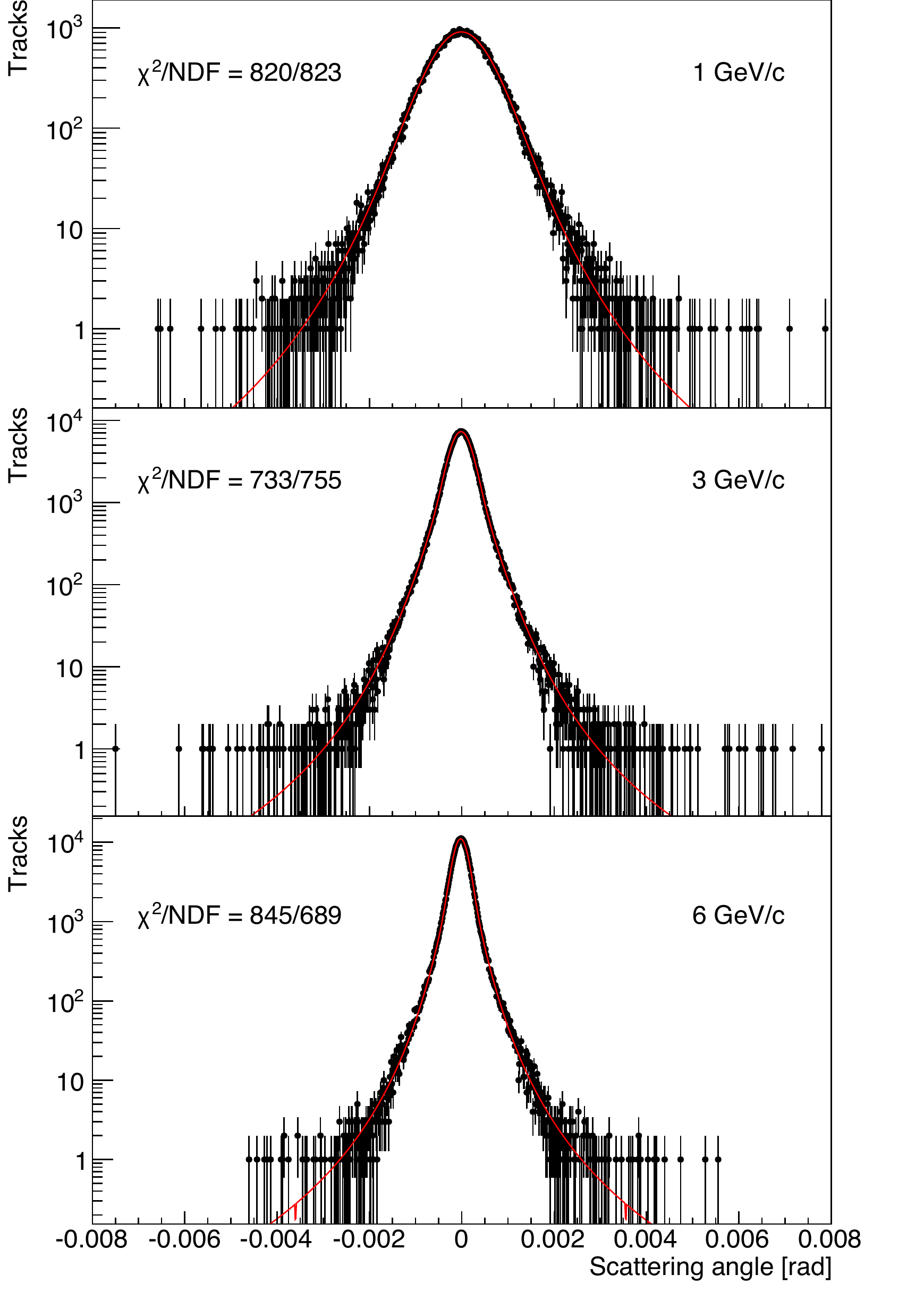}
	\caption{Horizontal scattering angle distribution for 1, 3 and 6~GeV/c electrons with 50~$\mu$m silicon target and an incidence angle to the beam of 15$^\circ$ in the device-under-test position. As fit function, a convolution of the shape obtained from a fit to the angular distribution without target and a Student's $t$-distribution is used as described in the text.}
	\label{fig:2_GeV-30_deg-1_sil}
\end{figure}

%\begin{figure}
	%\centering
		%\includegraphics[width=0.46\textwidth]{figures/3_GeV-45_deg-2_sil.pdf}
	%\caption{Horizontal scattering angle distribution for 3~GeV/c electrons with 100~$\mu$m silicon at 45$^\circ$ angle to the beam in the device-under-test position. The fit is a convolution of the shape obtained from a fit to the angular distribution with no silicon in the DUT position and a Student's $t$-distribution.}
	%\label{fig:3_GeV-45_deg-2_sil}
%\end{figure}

The telescope planes are aligned using reconstructed tracks in the configuration without silicon target using the EUTelescope software framework \cite{eutelescope}. Track residuals after alignment are below 2~$\mu$m. The distance of the target scattering plane to the third and fourth telescope planes is known to about 1~mm.

For the scattering analysis, tracks are reconstructed separately in the up- and downstream parts of the telescope and extrapolated to the silicon target plane. If an up- and a downstream track intersect within 150~$\mu$m on that plane and there are no matching ambiguities, the scattering angles $\theta$ between the tracks are calculated in both the horizontal and vertical projections.

The effect of multiple scattering in the telescope including the air surrounding the target together is larger than the scattering in the target. The measured distribution of planar scattering angles $f(\theta)$ can be described by convoluting the telescope scattering contributions with the scattering distribution in the target:
\begin{equation}
	f(\theta) = f_{\textnormal{telescope, upstream}} \otimes f_{\textnormal{target}} \otimes f_{\textnormal{telescope, downstream}}
\end{equation}
In order to determine the effect of the telescope, we first study the scattering angle distribution $f_{\textnormal{telescope}} = f_{\textnormal{telescope, upstream}} \otimes f_{\textnormal{telescope, downstream}}$ for datasets without silicon target in the beam. As an ansatz we use the sum of a Gaussian and a Student's $t$ distribution \cite{STUDENT}; the core of the scattering distribution is expected to be Gaussian and the Student's $t$ distribution can account for the large tails. Empirically we found that the measured distributions are well described by the this sum:
\begin{align}
		f_{\textnormal{telescope}}(\theta) \;=\;& N \cdot \left( (1-a) \cdot \frac{1}{\sigma_G \sqrt{2\pi}} e^{-\frac{(\theta-\mu)^2}{2 \sigma_G^2}} + \right. \notag \\
		&\left. a \cdot  \frac{\Gamma(\frac{\nu+1}{2})}{\sqrt{\nu\pi}\sigma\Gamma(\frac{\nu}{2})}\left(1 + \frac{(\theta-\mu)^2}{\nu \sigma^2} \right)^{-\frac{\nu+1}{2}}       \right).
		\label{eq:tel}
\end{align}
 A binned likelihood fit with six free parameters, namely overall normalization $N$, relative fraction $a$ of the Student's $t$ distribution, a common mean $\mu$, the width  of the Gaussian $\sigma_G$ and the width $\sigma$ and tail parameter $\nu$ of the $t$ distribution is used. For $\nu \rightarrow \infty$, the Student's $t$ distribution turns into a Gaussian, whereas for $\nu \rightarrow 1$, the tails get more pronounced. At $\nu = 1$, a Lorentzian distribution is obtained. We obtain good fits at all electron momenta. 
Figure~\ref{fig:nosilicon} shows the fitted horizontal scattering angle distributions at 1, 3 and 6~GeV/$c$ electron momentum. The fits for the horizontal and vertical scattering angles give results that are compatible within statistical uncertainties, thus reassuring us that there are no large residual effects of telescope misalignment or acceptance.

The data with the silicon target in the beam are fitted using a binned likelihood function based on the convolution of a Student's $t$ distribution, representing the contribution by the target, and the shape of the scattering angle distribution of the telescope and air as given in equation~\ref{eq:tel}:
%\footnote{We found that fitting a convolution is numerically much more stable than performing a deconvolution.}; 
\begin{equation}
	f(\theta) = N \cdot \int f_{\textnormal{telescope}}(\theta-\tau) \cdot \frac{\Gamma(\frac{\nu+1}{2})}{\sqrt{\nu\pi}\sigma\Gamma(\frac{\nu}{2})}\left(1 + \frac{(\tau)^2}{\nu \sigma^2} \right)^{-\frac{\nu+1}{2}} \mathrm{d}\tau.
\end{equation}
The free parameters in this fit are the overall normalization $N$ and the width $\sigma$ and tail parameter $\nu$ of the $t$ distribution. Again we obtain good fits, see Figure~\ref{fig:2_GeV-30_deg-1_sil}.% and \ref{fig:3_GeV-45_deg-2_sil}.

The fits are performed for the horizontal and vertical scattering angles separately; the results are consistent within uncertainties. All the figures shown in the following are based on a combination of the results from the two projections. All fit results and their statistical uncertainties are listed in Table~\ref{tab:MeasuredAndSimulatedValues}.

%\begin{figure}
	%\centering
		%\includegraphics[width=0.45\textwidth]{figures/sigma_momentum.pdf}
	%\caption{Fitted $\sigma$ of the Student $t$ distribution versus electron momentum for varying silicon thickness. The data points are slightly offset from their horizontal positions at multiples of 1~GeV/$c$ for better visibility.}
	%\label{fig:sigma_momentum}
%\end{figure}

\begin{figure}[t!]
	\centering
		\includegraphics[width=0.49\textwidth]{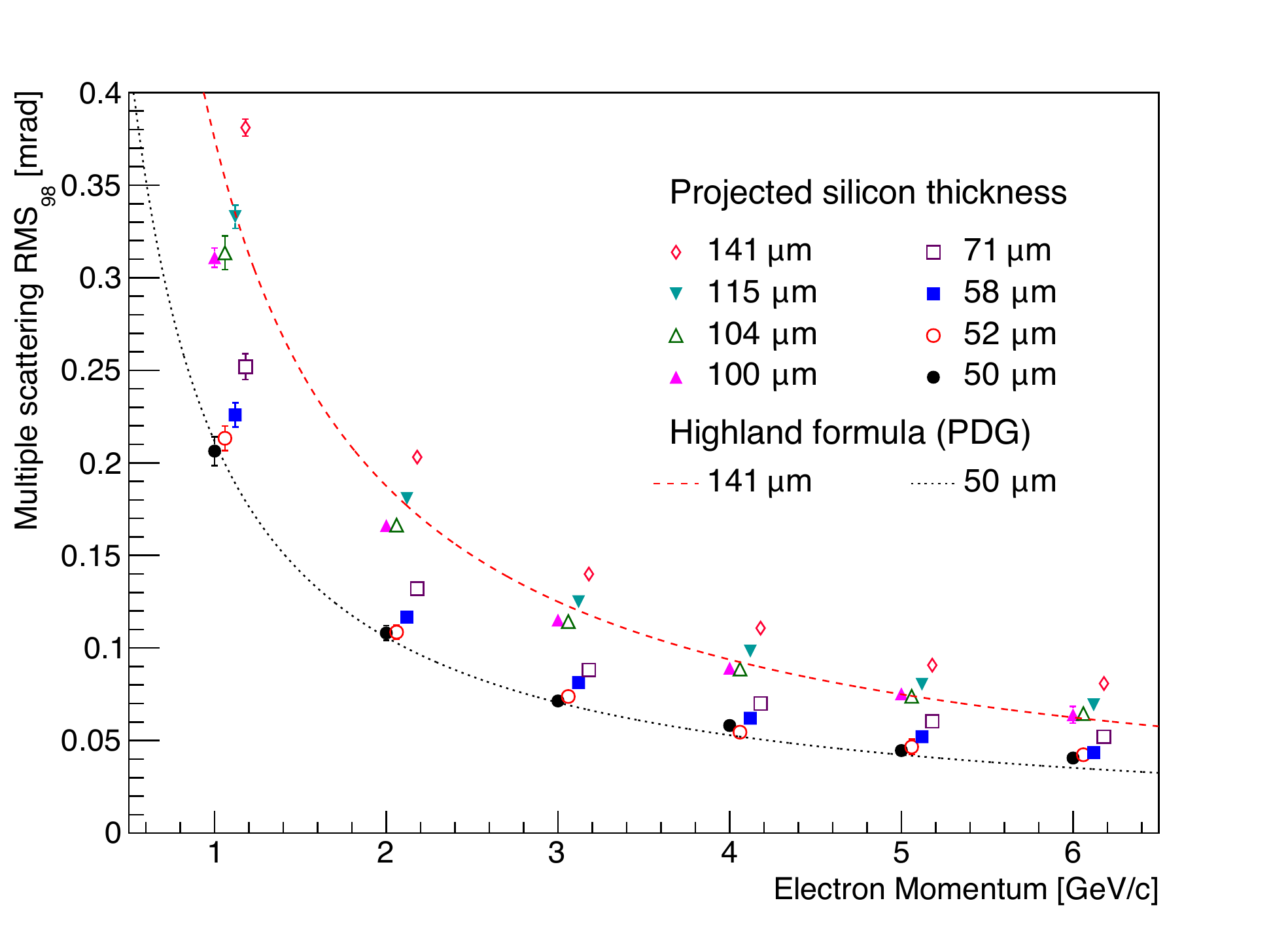}
		\includegraphics[width=0.49\textwidth]{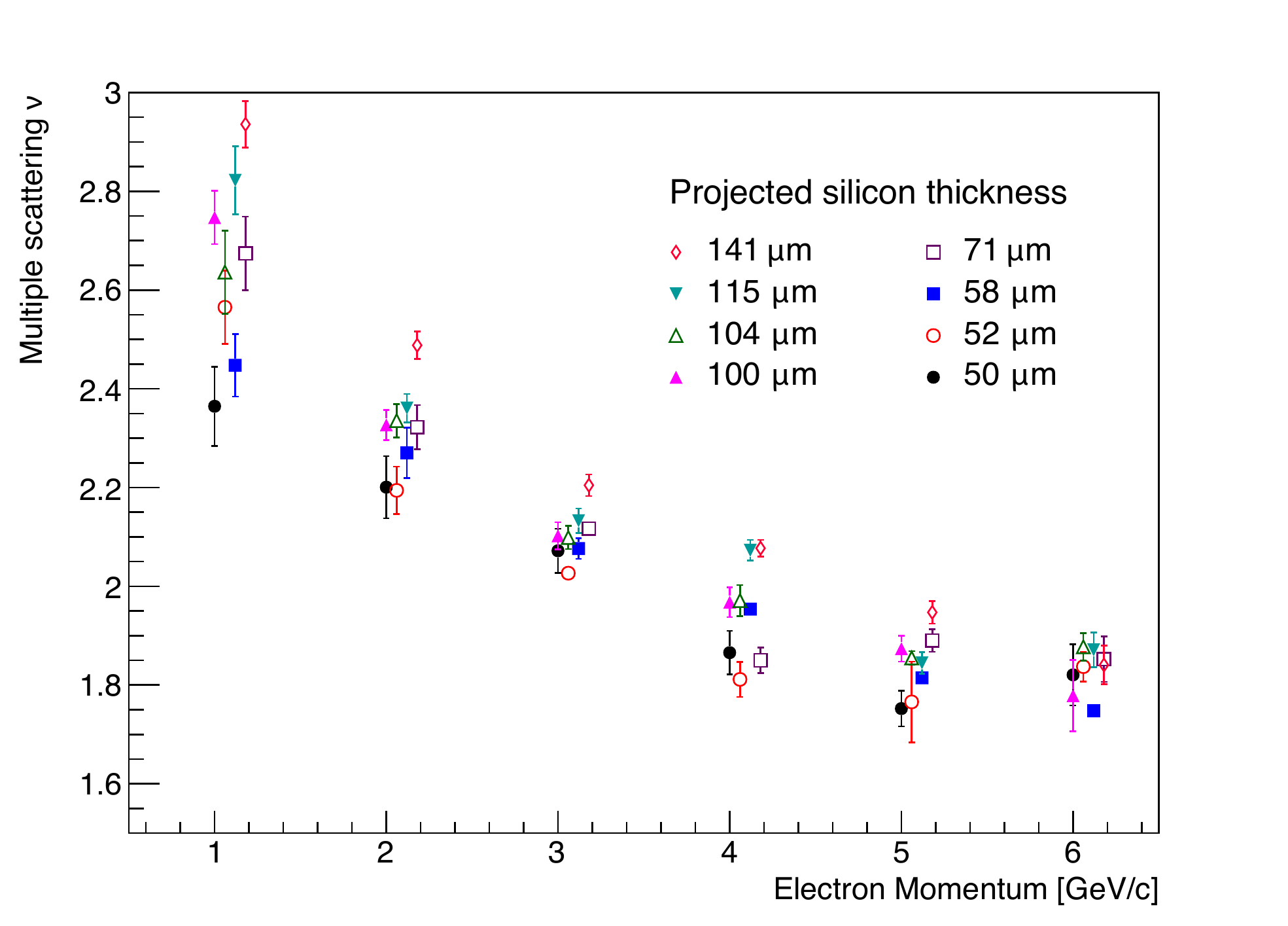}
	\caption{$RMS$ of the central 98\% of the fitted Student's $t$ distribution versus electron momentum for varying silicon target thickness compared to the Highland-parametrisation (left). Fitted tail parameter $\nu$ of the Student $t$ distribution versus electron momentum for varying silicon target thickness (right). The data points are slightly offset from their horizontal positions at multiples of 1\,GeV/$c$ for better visibility. The error bars represent the 1\,$\sigma$ uncertainty of the fit.
	%; where they cannot be discerned, they are smaller than the markers. 
	Smaller $\nu$ values correspond to larger tail fractions.}
	\label{fig:data_momentum}
\end{figure}

%\begin{figure}
%	\centering
%		
%	\caption{ The data points are slightly offset from their horizontal positions at multiples of 1~GeV/$c$ for better visibility. The error bars represent the 1~$\sigma$ uncertainty of the fit; where they cannot be discerned, they are smaller than the markers.}
%	\label{fig:rms_momentum}
%\end{figure}

%\begin{figure}
	%\centering
		%\includegraphics[width=0.45\textwidth]{figures/sigma_thickness.pdf}
	%\caption{Fitted $\sigma$ of the Student $t$ distribution versus silicon thickness for varying electron momenta.}
	%\label{fig:sigma_thickness}
%\end{figure}

\begin{figure}[t!]
	\centering
		\includegraphics[width=0.49\textwidth]{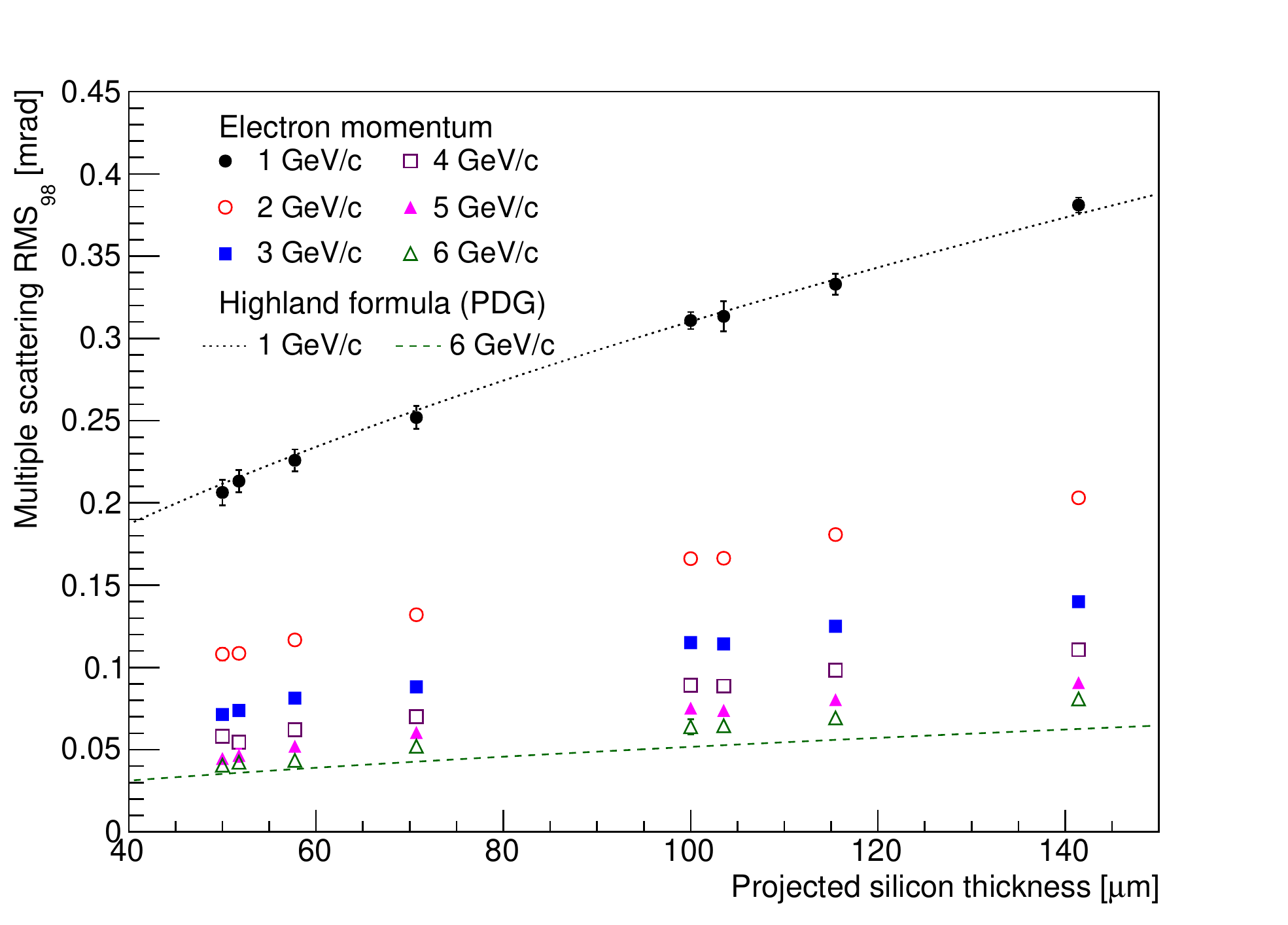}
		\includegraphics[width=0.49\textwidth]{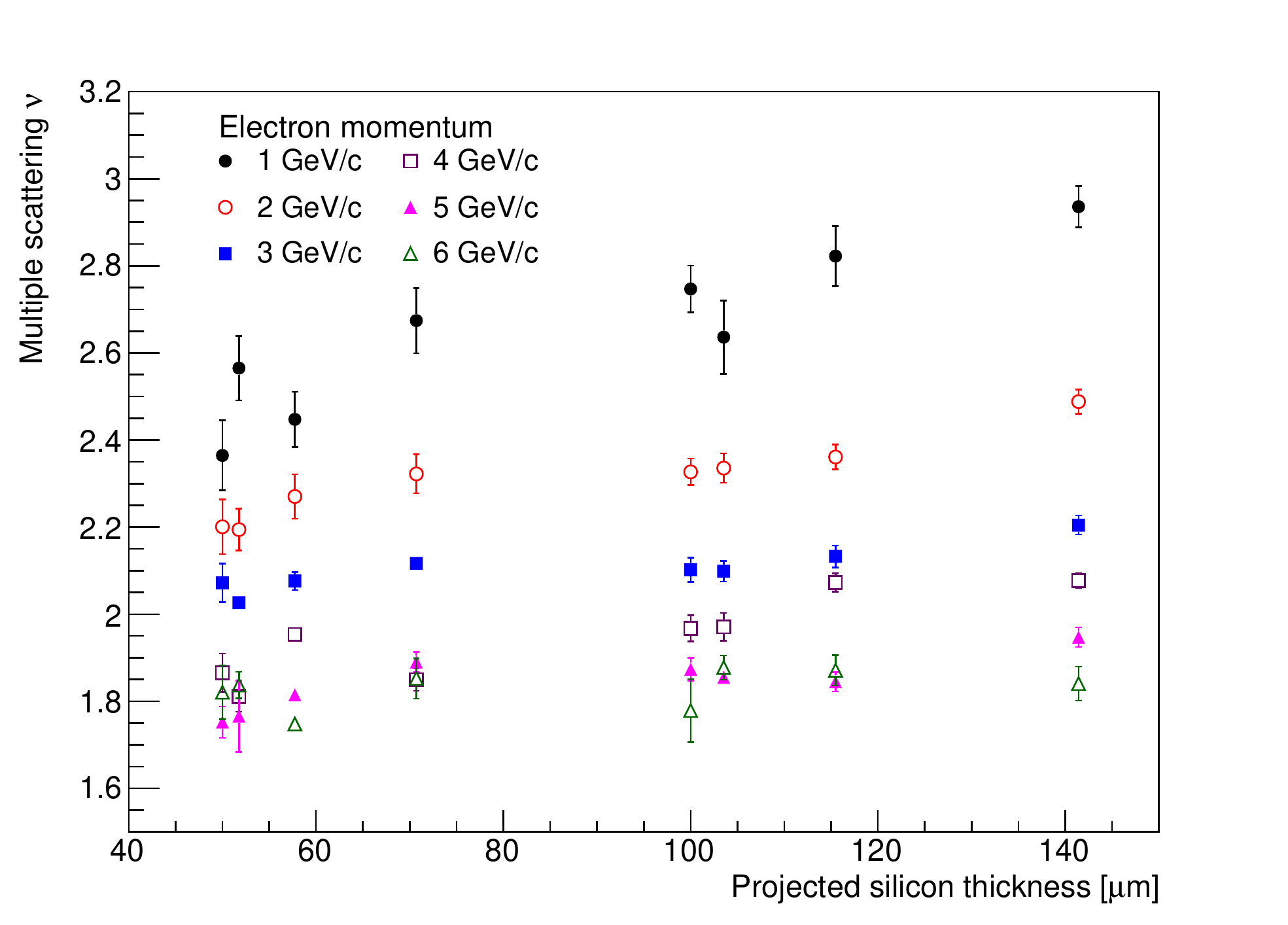}
	\caption{$RMS$ of the central 98\% of the fitted Student's $t$ distribution versus silicon target thickness for varying electron momenta compared to the Highland-parametrisation (left). Fitted tail parameter $\nu$ of the Student $t$ distribution versus silicon target thickness for varying electron momenta (right). The error bars represent the 1\,$\sigma$ uncertainty of the fit.%; where they cannot be discerned, they are smaller than the markers.
	}
	\label{fig:data_thickness}
\end{figure}

%\begin{figure}
%	\centering
		
%	\caption{RMS of the central 98\% of the fitted Student $t$ distribution versus silicon thickness for varying electron momenta compared to the Highland-parametrisation. The error bars represent the 1~$\sigma$ uncertainty of the fit; where they cannot be discerned, they are smaller than the markers.}
%	\label{fig:rms_thickness}
%\end{figure}

%\begin{figure}
	%\centering
		%\includegraphics[width=0.49\textwidth]{figures/sim_data_rms_momentum_plustheory_50um.pdf}
	%\caption{Comparison of the RMS of the central 98\% of the fitted Student $t$ distribution versus momentum for various scattering models in Geant4 with our data obtained with a 50~$\mu$m silicon wafer perpendicular to the beam. The Highland parametrisation is also shown for reference.}
	%\label{fig:sim_data_rms_momentum_50um}
%\end{figure}

The $RMS_{98}$ as well as the tail parameter $\nu$ of the distributions are shown as a function of electron momentum in Figure~\ref{fig:data_momentum} and effective thickness in Figure~\ref{fig:data_thickness}. As can be seen in the left panels of Figures~\ref{fig:data_momentum} and \ref{fig:data_thickness}, the $RMS$ of the core of the scattering distributions is described by the Highland formula within the 10\% uncertainty quoted \cite{PDG2012}. The amount of tails increases with momenta, see the right panel of Figure~\ref{fig:data_momentum}.  This is expected, as higher momentum electrons get closer to the nuclei of the scatterer and thus see a less screened nuclear potential leading to larger deflections. The tail fraction also slightly decreases with thickness, see the right panel of Figure~\ref{fig:data_thickness}; this seems to indicate that for the thin scatterers used here, the statistical approach to multiple scattering starts to break down as individual large angle scattering events become important.

%For scattering models in a simulation such as Geant4, adequately describing this shape of the scattering distribution is an additional challenge. 
 
\section{Comparison with Simulation Models}

In order to compare the results with multiple scattering models, we simulate one million electron tracks propagating through 50 and 100~$\mu m$ of silicon at incident angles of 0$^\circ$, 15$^\circ$, 30$^\circ$ and 45$^\circ$. The simulated scattering distributions are fitted with a Student's $t$ distribution. The following models in Geant4 \cite{Agostinelli2003250, Allison:2006ve} are tested:

\begin{itemize}
	\item \emph{Single scattering}: Electrons are propagated from one Coulomb scattering to the next. This procedure should give the most accurate results, assuming the data on scattering lengths and the screened nuclear potential are adequate. However, for any simulation of moderately complex set-ups involving solids, this approach is too computing-intensive.
	\item \emph{Urb\'an}: The standard multiple scattering model in Geant4, based on the theoretical work of Lewis \cite{Lewis:1950zz}. A recently re-tuned model is available in Geant4 version 10.0~\footnote{The older model is taken from Geant4 version 9.6 patch 2.}.
	\item \emph{Goudsmit-Saunderson}\cite{Goudsmit:1940zza, Goudsmit:1940zz}: The model produces a purely Gaussian distribution for our set-up, the $\nu$ parameter thus is fitted at very large values above 100 and therefore not shown in the following figures.
	\item \emph{Our model}: A model drawing scattering angles from a Student's $t$ distribution with parameters tuned to our data, details are described in section \ref{sec:Model}.
\end{itemize}

\begin{figure}[t!]
	\centering
	\includegraphics[width=0.49\textwidth]{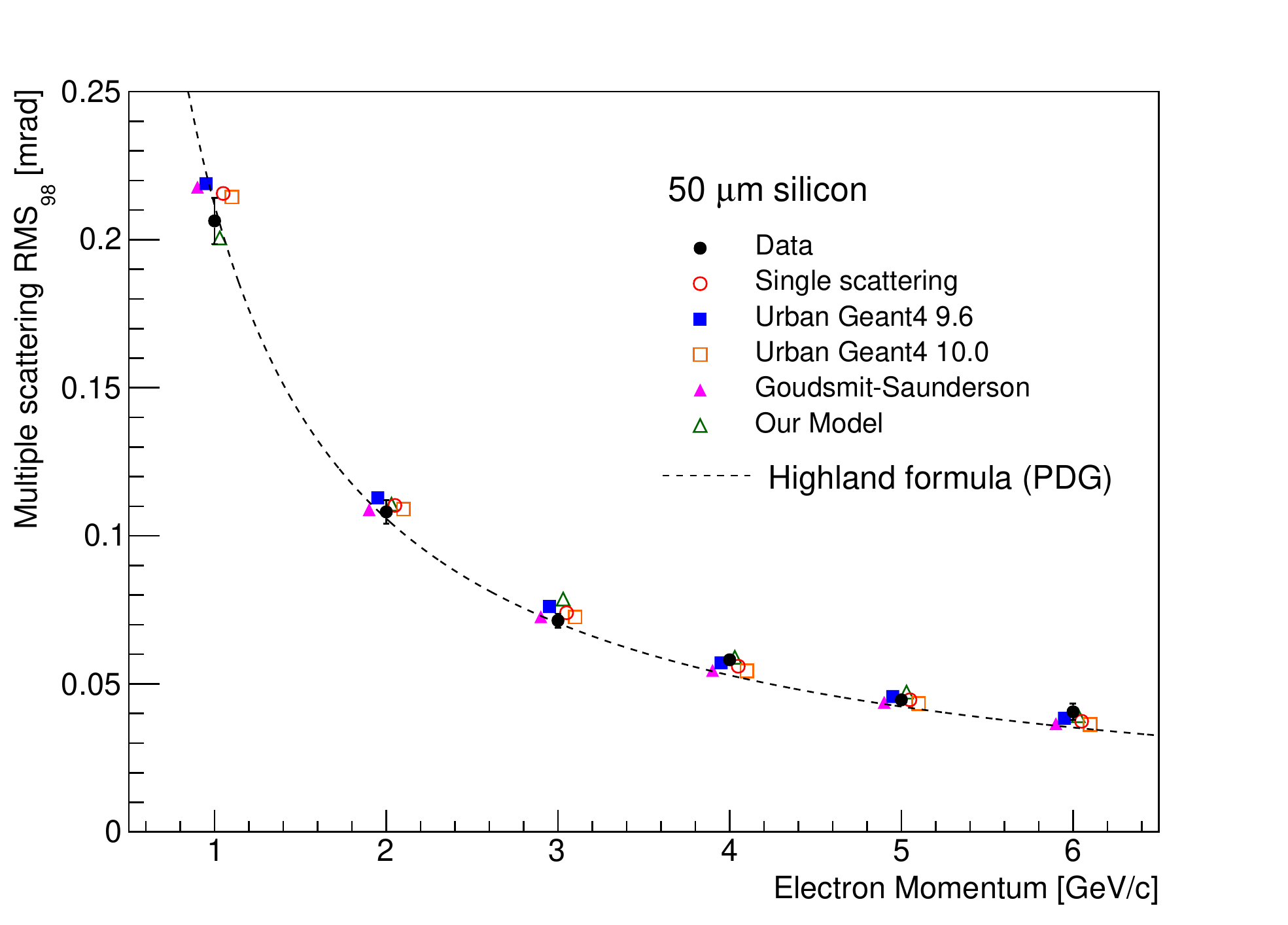}
		\includegraphics[width=0.49\textwidth]{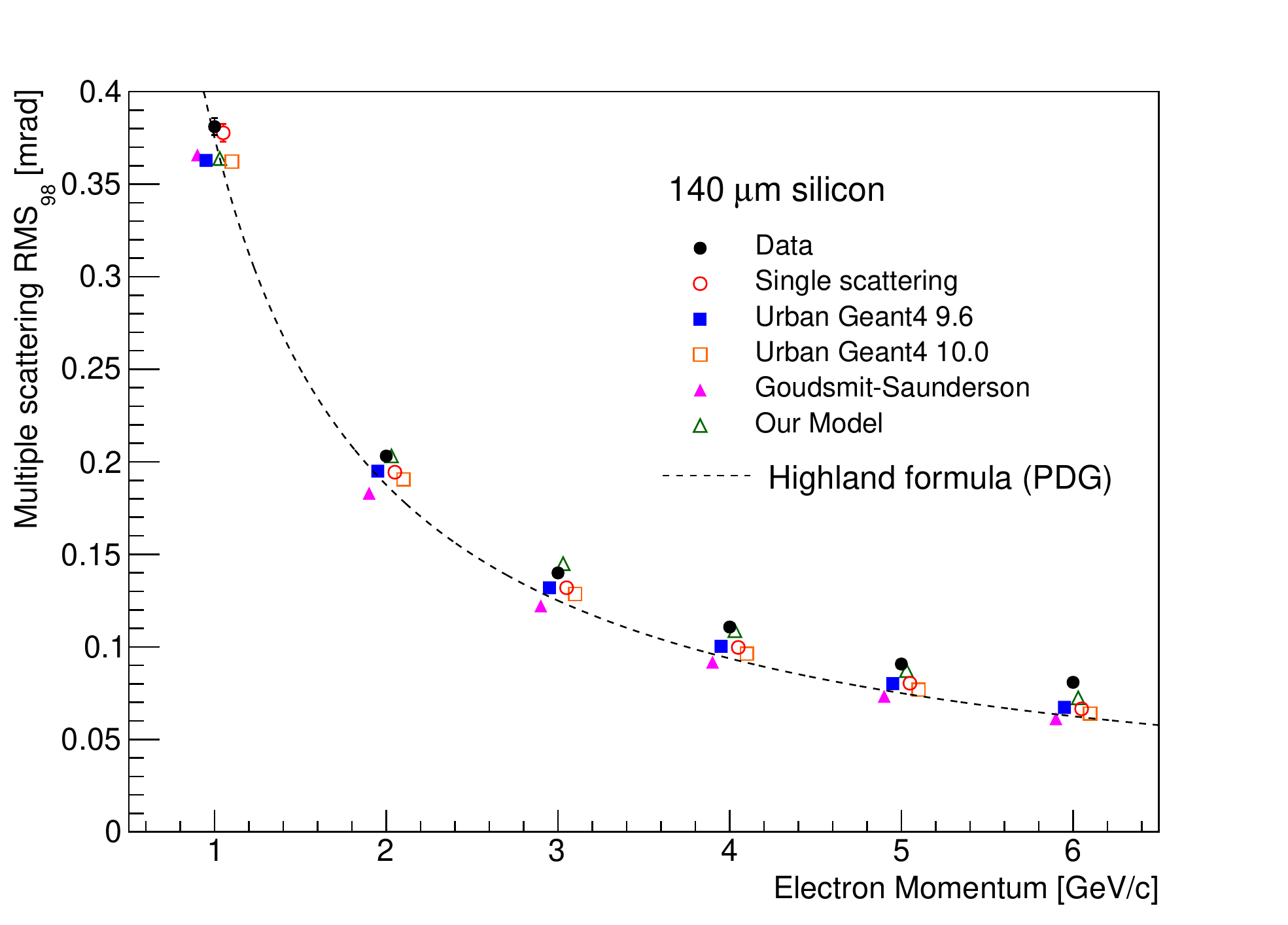}
	\caption{Comparison of the $RMS$ of the central 98\% of the fitted Student's $t$ distribution versus momentum for various scattering models in Geant4 with our data obtained with a 50~$\mu$m silicon target perpendicular to the beam (left) and a 100~$\mu$m silicon target tilted by 45$^\circ$ (right). The data points are slightly offset from their horizontal positions at multiples of 1~GeV/$c$ for better visibility. The Highland parametrisation is also shown for reference. The error bars represent the statistical uncertainty of the fit.}
	\label{fig:sim_data_rms_momentum}
\end{figure}

\begin{figure}[t!]
	\centering
		\includegraphics[width=0.49\textwidth]{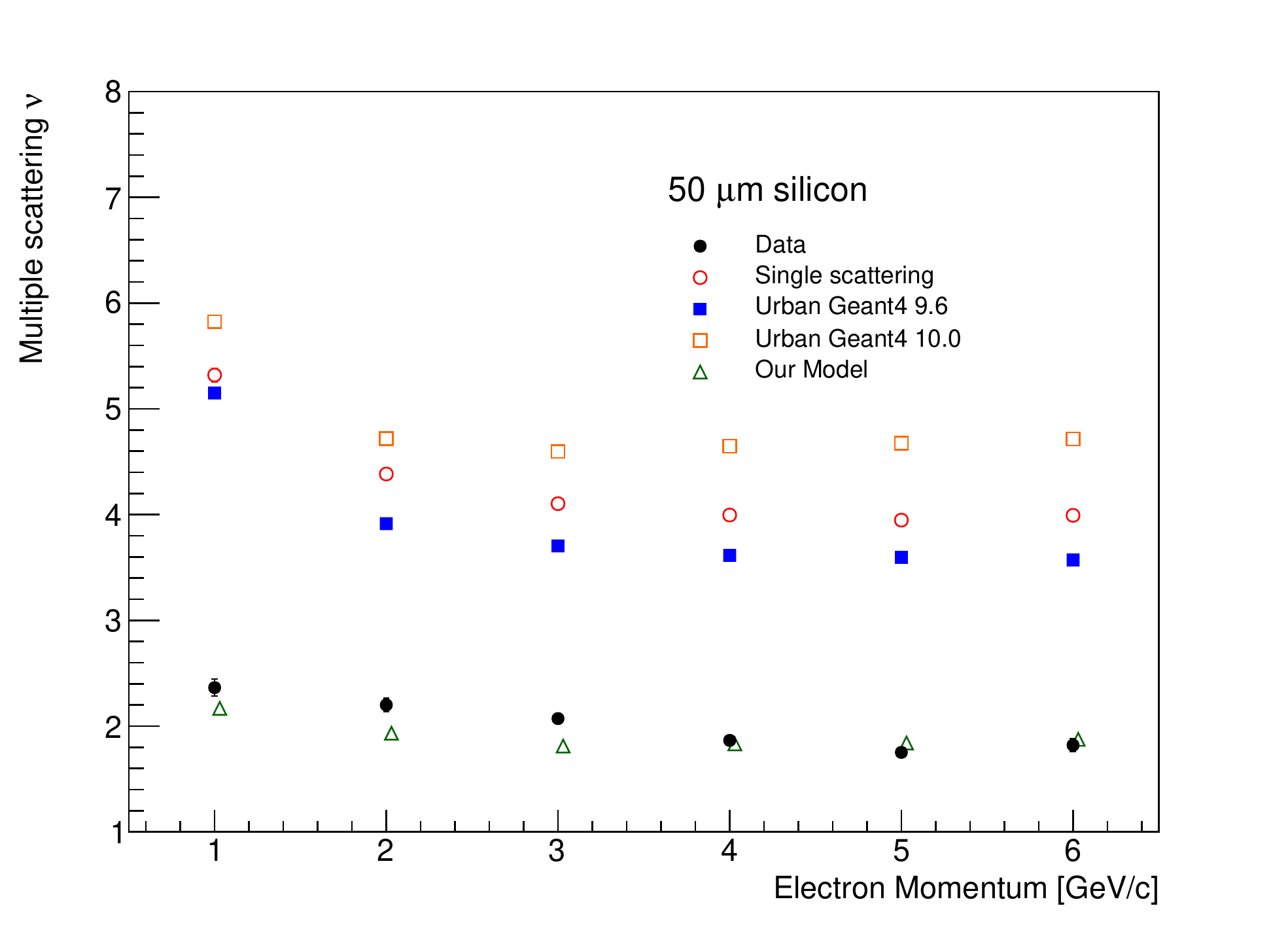}
		\includegraphics[width=0.49\textwidth]{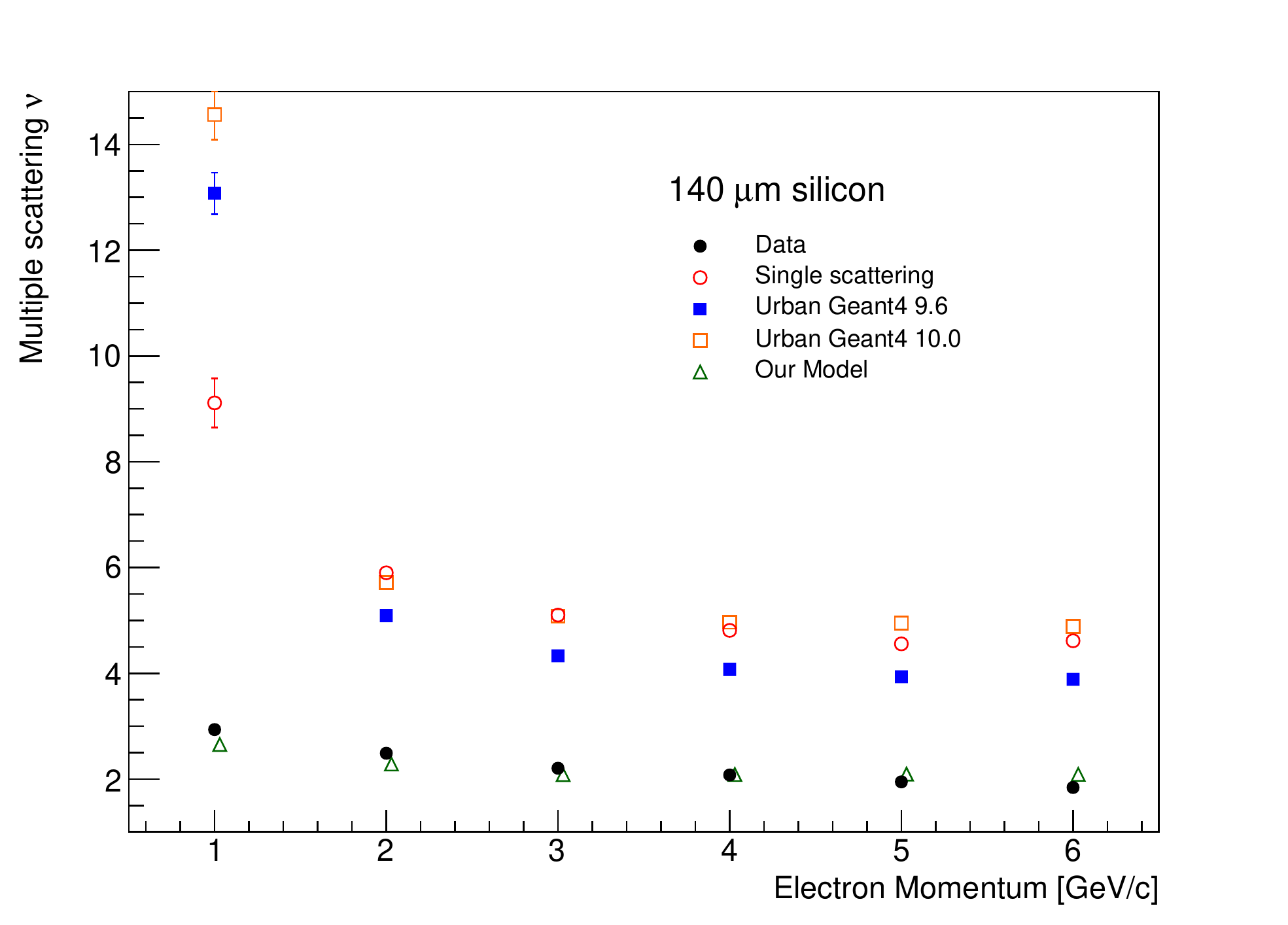}
	\caption{Comparison of the Student's $t$ tail parameter $\nu$ versus momentum for various scattering models in Geant4 with our data obtained with a 50~$\mu$m silicon target perpendicular to the beam (left) and a 100~$\mu$m silicon target tilted by 45$^\circ$ (right). The error bars represent the statistical uncertainty of the fit.}
	\label{fig:sim_data_nu_momentum}
\end{figure}

%%%%%%%%%%%%%%%%%%%%%%%%%%

\begin{figure}[t!]
	\centering
		\includegraphics[width=0.49\textwidth]{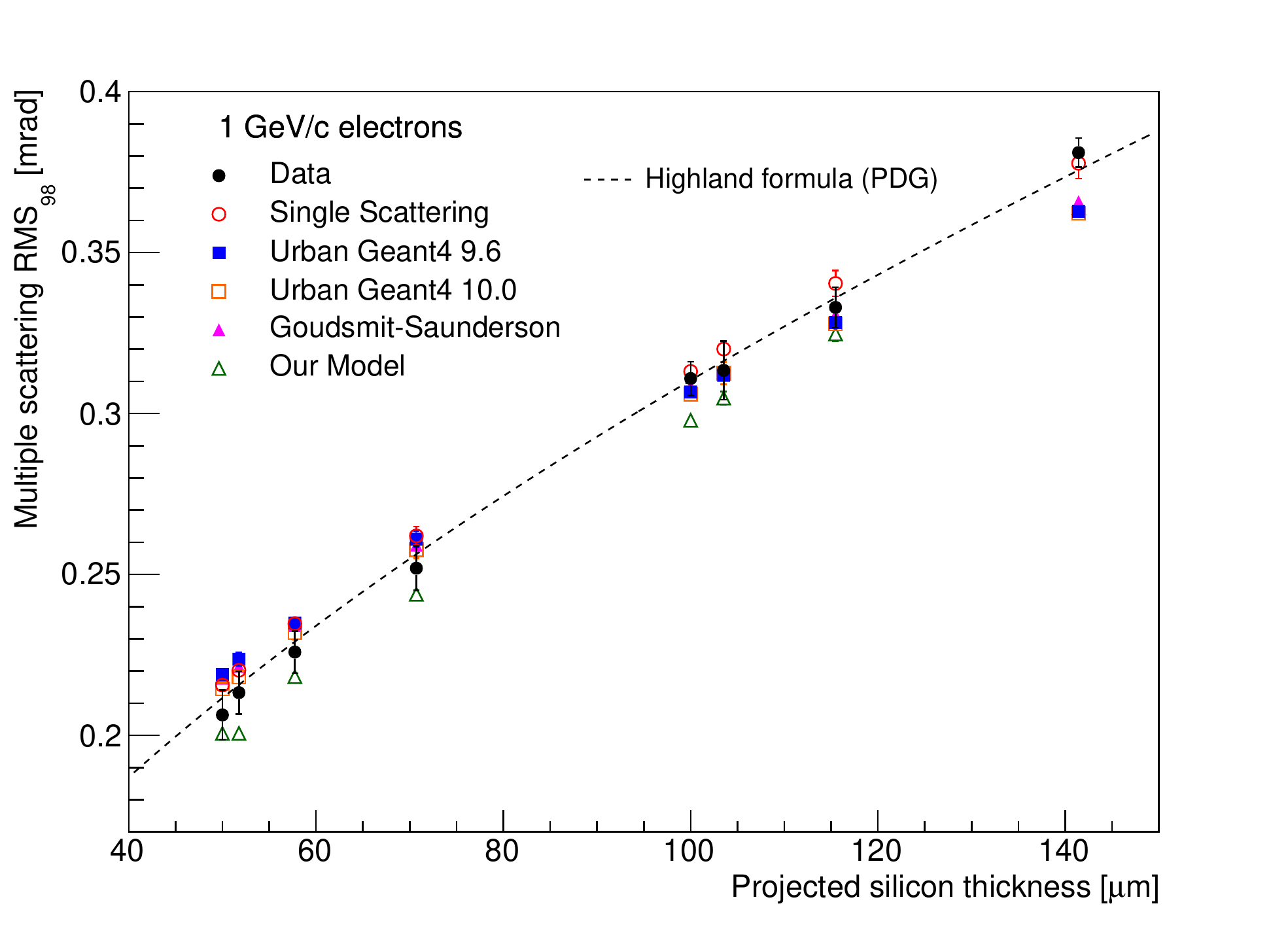}
		\includegraphics[width=0.49\textwidth]{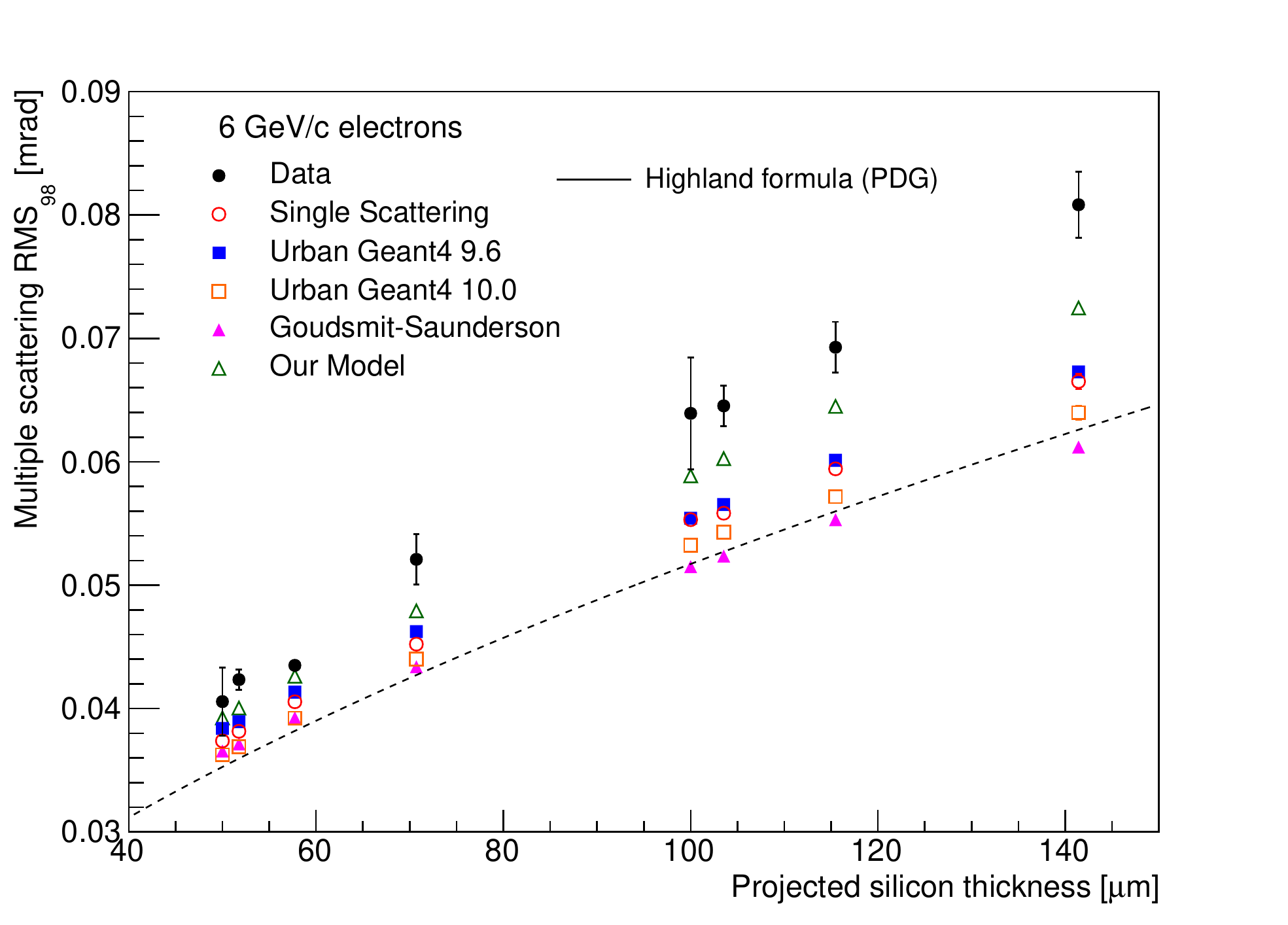}
	\caption{Comparison of the $RMS$ of the central 98\% of the fitted Student's $t$ distribution versus projected silicon target thickness for various scattering models in Geant4 with our data obtained at 1~GeV/$c$ (left) and 6~GeV/$c$ (right) electron momentum. The error bars represent the statistical uncertainty of the fit. The Highland parametrisation is also shown for reference.}
	\label{fig:sim_data_rms_thickness}
\end{figure}

\begin{figure}[t!]
	\centering
		\includegraphics[width=0.49\textwidth]{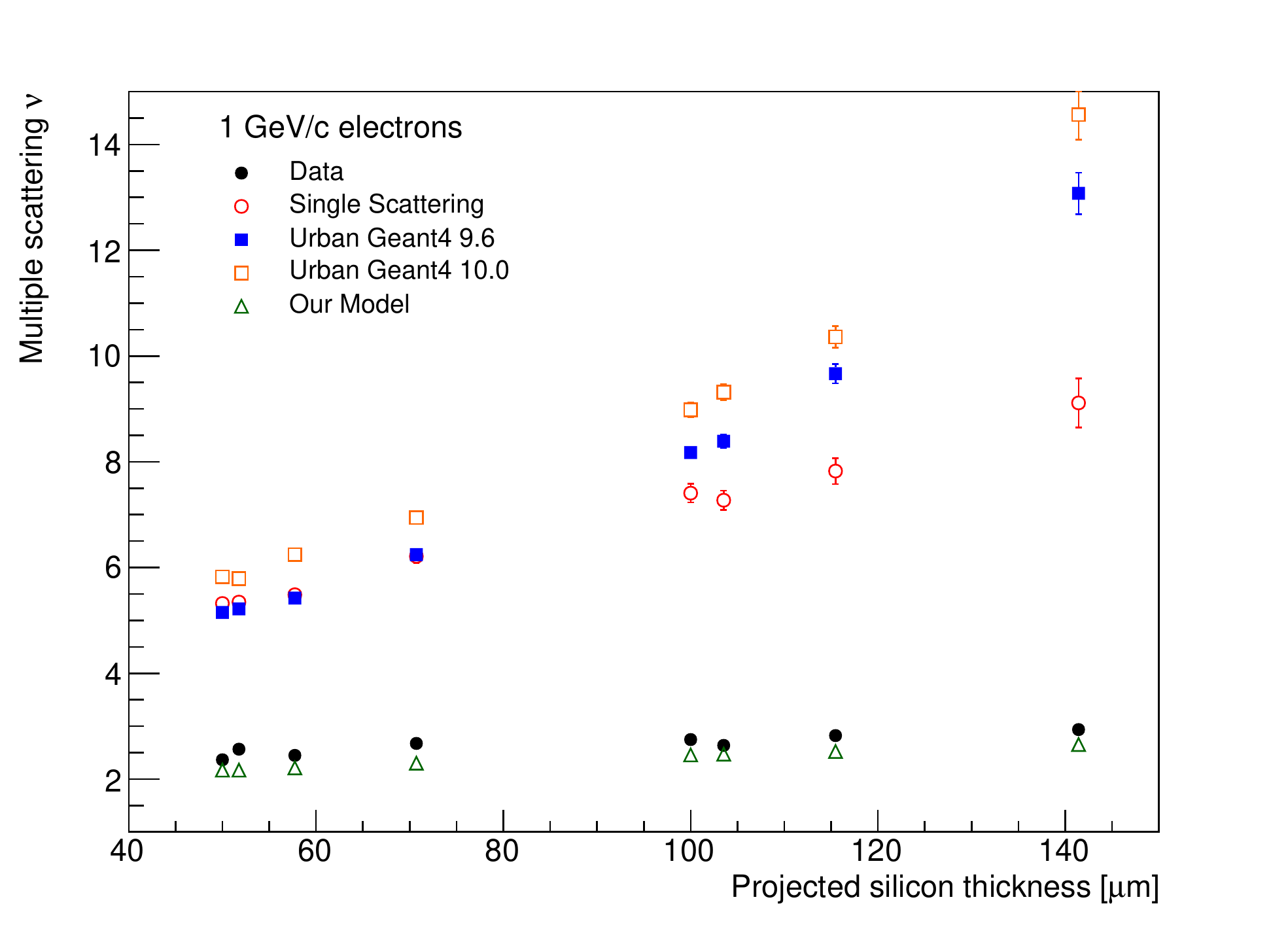}
		\includegraphics[width=0.49\textwidth]{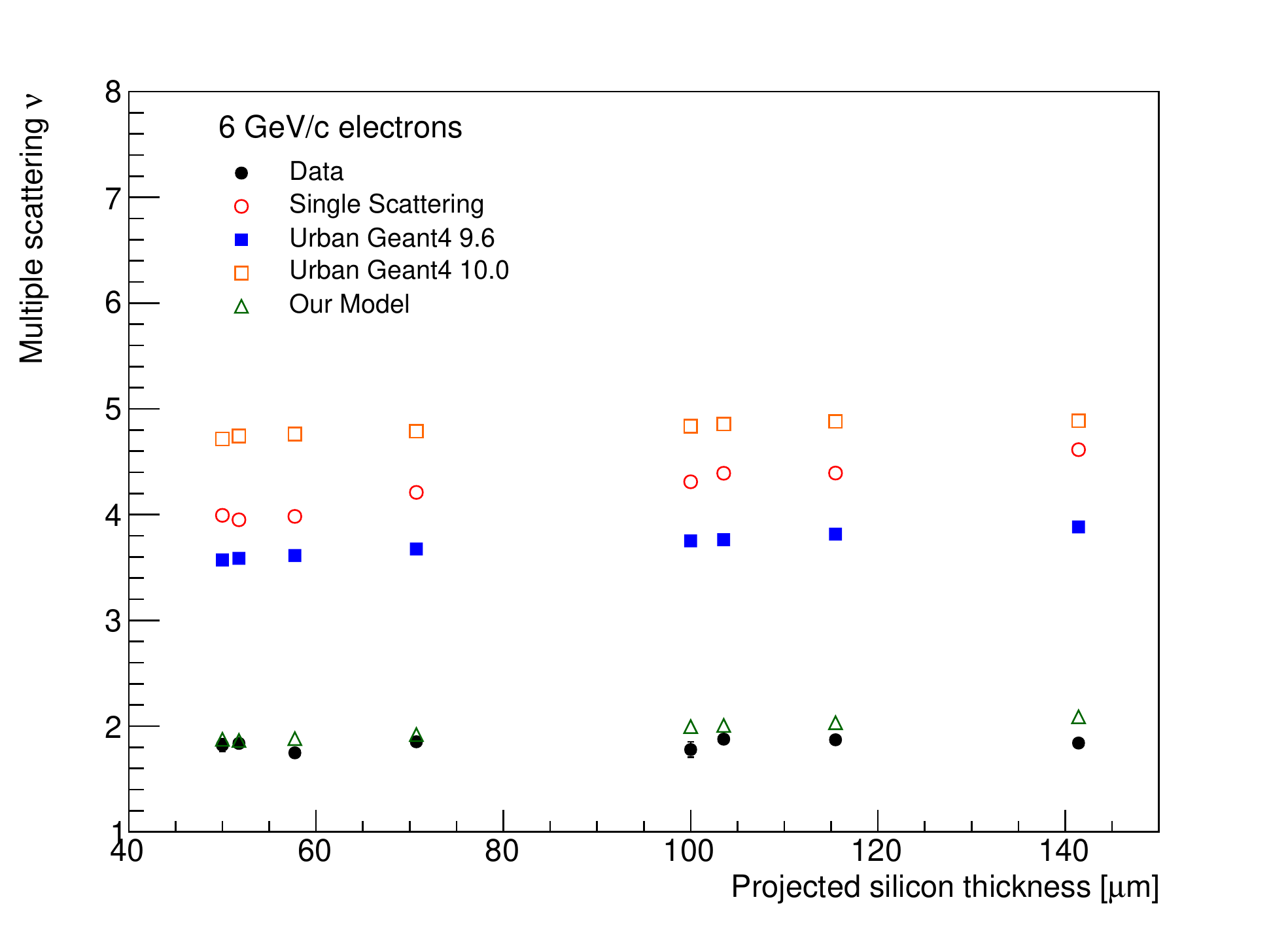}
	\caption{Comparison of the Student's $t$ tail parameter $\nu$ versus projected silicon target thickness for various scattering models in Geant4 with our data obtained at 1~GeV/$c$ (left) and 6~GeV/$c$ (right) electron momentum. The error bars represent the statistical uncertainty of the fit.}
	\label{fig:sim_data_nu_thickness}
\end{figure}

Comparisons of the models with our data as a function of electron momentum are presented in Figures \ref{fig:sim_data_rms_momentum} and \ref{fig:sim_data_nu_momentum} and as a function of effective thickness in Figures~\ref{fig:sim_data_rms_thickness} and \ref{fig:sim_data_nu_thickness}. The $RMS_{98}$ of the distributions is well described by all models, including the Highland parametrisation; however the data show a markedly higher tail fraction (and a correspondingly narrower core) than all the models. The difference is more pronounced at low momenta.

\section{A New Multiple Coulomb Scattering Model}
\label{sec:Model}

Building on the success of the various models in describing the $RMS_{98}$ of the scattering distribution, we built a new model for the Geant4 framework with a better description of the shape of the distribution. It is based on the Urb\'an model and reuses the code for the calculation of the $RMS$ of the scattering angle (essentially the Highland parametrisation), but instead of using different parametrisations for core and tail of the distribution, it draws all angles from a Student's $t$ distribution. The tail parameter $\nu$ of the distribution is obtained from an empirical fit to our data of the form
\begin{equation}
	\nu(p,d)_{\textnormal{fit}} = A + B \cdot \frac{1}{p-D} + C \cdot d 
\end{equation}
where $p$ is the electron momentum in GeV/$c$ and $d$ the silicon thickness in radiation lengths. $A$, $B$ $C$ and $D$ are the fit parameters; the numerical values are shown in table \ref{tab:NumericalValuesOfTheParametersInTheStudentTScatteringModel}. 

As Geant4 sometimes splits the tracking step through the thin silicon in two (e.g.~due to emission of a $\delta$-electron), the approach described above invariably produces too much tails. The input angular distribution in Geant4 is thus different from the scattering distribution in the simulation output. It turns out that forcing the $\nu$ parameter to be at least two, 
\begin{equation}
	\nu_{\textnormal{Geant}} = \max(\nu(p,d)_{\textnormal{fit}}, 2),
\end{equation}
leads to a much improved description of the scattering distribution shape of our data. A comparison of our model to the data and existing Geant4 models can be seen in Figures \ref{fig:sim_data_rms_momentum} to \ref{fig:sim_data_nu_thickness}. The small differences in the $RMS_{98}$ between our model and the Urb\'an model are partly due to the multiple step effect and partly due to the differences between the full $RMS$ and the $RMS_{98}$.

\begin{table}
	\centering
		\begin{tabular}{lrr}
			\hline
			Parameter & Value from fit & Uncertainty from fit \\
			\hline
				A				& 1.10					 & 0.07\\
				B       & 4.36           & 0.65\\
				C       & 1.90           & 0.17\\
				D				& - 2.04				 & 0.37\\
			\hline
		\end{tabular}
	\caption{Numerical values of the parameters in the Student $t$ scattering model.}
	\label{tab:NumericalValuesOfTheParametersInTheStudentTScatteringModel}
\end{table}

\section{Conclusion}

We have measured multiple Coulomb scattering of 1-6~GeV/$c$ electrons in a 50-141~$\mu$m thin silicon target. We found a good description of the scattering distribution $RMS_{98}$ width in data by models implemented in Geant4, but large differences in the tail fraction of the distribution. A newly developed model based on drawing scattering angles from a Student's $t$ distribution with parameters obtained from our data gives a greatly improved shape description for ultra-thin silicon trackers as presently used by many experiments.

\acknowledgments

The support of the Deutsches Elektronensynchrotron (DESY) providing the test beam and the related infrastructure made this measurement possible.
We would especially like to thank the EUDET telescope group at DESY, in particular \mbox{I.-M.~Gregor}, I.~Rubinsky and M.~Stanitzki for their valuable support of this test beam measurement.

N.~Berger would like to thank the \emph{Deutsche Forschungsgemeinschaft} for support through an Emmy Noether grant. M.~Kiehn acknowledges support by the \emph{International Max Planck Research School on Precision Tests of Fundametnal symmetries} and O.~Kovalenko was supported by the \emph{Heidelberg Graduate School for Fundamental Physics}.

\begin{table*}
	\centering
	\tiny
		\begin{tabular}{rr|rrr|rrr|rrr}
\hline
   &  & \multicolumn{3}{c|}{Data} &  \multicolumn{3}{c|}{Urban} & \multicolumn{3}{c}{Our Model}\\
   & & \multicolumn{3}{c|}{}     & \multicolumn{3}{c|}{Geant4 10.0} &\multicolumn{3}{c}{}\\
 $p$ & $d$ & $\sigma$ & $\nu$ & $RMS_{98}$& $\sigma$ & $\nu$ & $RMS_{98}$& $\sigma$ & $\nu$ & $RMS_{98}$\\
 GeV/$c$ & $\mu$m& 0.1 mrad &   & 0.1 mrad& 0.1 mrad &   & 0.1 mrad& 0.1 mrad &   & 0.1 mrad\\
\hline
1 & 50 & $1.404 \pm 0.045 $  & $2.36 \pm 0.08 $  & $2.064 \pm 0.078 $  & $1.962 \pm 0.002 $  & $5.83 \pm 0.04 $  & $2.144 \pm 0.003 $  & $1.292 \pm 0.002 $  & $2.17 \pm 0.01 $  & $2.006 \pm 0.004 $ \\
 & 52 & $1.513 \pm 0.039 $  & $2.57 \pm 0.07 $  & $2.133 \pm 0.067 $  & $1.995 \pm 0.003 $  & $5.79 \pm 0.04 $  & $2.181 \pm 0.003 $  & $1.292 \pm 0.002 $  & $2.17 \pm 0.01 $  & $2.006 \pm 0.004 $ \\
 & 58 & $1.567 \pm 0.036 $  & $2.45 \pm 0.06 $  & $2.259 \pm 0.066 $  & $2.132 \pm 0.003 $  & $6.24 \pm 0.05 $  & $2.319 \pm 0.023 $  & $1.420 \pm 0.002 $  & $2.21 \pm 0.01 $  & $2.182 \pm 0.004 $ \\
 & 71 & $1.824 \pm 0.039 $  & $2.67 \pm 0.07 $  & $2.519 \pm 0.069 $  & $2.406 \pm 0.003 $  & $6.95 \pm 0.07 $  & $2.576 \pm 0.026 $  & $1.630 \pm 0.002 $  & $2.30 \pm 0.01 $  & $2.438 \pm 0.004 $ \\
 & 100 & $2.291 \pm 0.030 $  & $2.75 \pm 0.05 $  & $3.109 \pm 0.052 $  & $2.951 \pm 0.004 $  & $8.98 \pm 0.14 $  & $3.060 \pm 0.005 $  & $2.069 \pm 0.003 $  & $2.46 \pm 0.01 $  & $2.979 \pm 0.005 $ \\
 & 104 & $2.263 \pm 0.050 $  & $2.64 \pm 0.08 $  & $3.134 \pm 0.091 $  & $3.018 \pm 0.004 $  & $9.32 \pm 0.15 $  & $3.125 \pm 0.034 $  & $2.119 \pm 0.003 $  & $2.47 \pm 0.01 $  & $3.048 \pm 0.021 $ \\
 & 115 & $2.482 \pm 0.038 $  & $2.82 \pm 0.07 $  & $3.330 \pm 0.063 $  & $3.211 \pm 0.004 $  & $10.36 \pm 0.20 $  & $3.278 \pm 0.005 $  & $2.281 \pm 0.003 $  & $2.52 \pm 0.01 $  & $3.247 \pm 0.022 $ \\
 & 141 & $2.879 \pm 0.022 $  & $2.94 \pm 0.05 $  & $3.811 \pm 0.045 $  & $3.622 \pm 0.005 $  & $14.57 \pm 0.47 $  & $3.622 \pm 0.006 $  & $2.631 \pm 0.004 $  & $2.65 \pm 0.01 $  & $3.639 \pm 0.006 $ \\
\hline
2 & 50 & $0.703 \pm 0.022 $  & $2.20 \pm 0.06 $  & $1.081 \pm 0.040 $  & $0.956 \pm 0.001 $  & $4.72 \pm 0.02 $  & $1.090 \pm 0.001 $  & $0.652 \pm 0.001 $  & $1.93 \pm 0.01 $  & $1.107 \pm 0.007 $ \\
 & 52 & $0.705 \pm 0.021 $  & $2.19 \pm 0.05 $  & $1.086 \pm 0.037 $  & $0.975 \pm 0.001 $  & $4.72 \pm 0.02 $  & $1.113 \pm 0.001 $  & $0.668 \pm 0.001 $  & $1.94 \pm 0.01 $  & $1.127 \pm 0.002 $ \\
 & 58 & $0.773 \pm 0.018 $  & $2.27 \pm 0.05 $  & $1.167 \pm 0.034 $  & $1.038 \pm 0.001 $  & $4.77 \pm 0.02 $  & $1.183 \pm 0.002 $  & $0.720 \pm 0.001 $  & $1.99 \pm 0.01 $  & $1.194 \pm 0.007 $ \\
 & 71 & $0.889 \pm 0.017 $  & $2.32 \pm 0.04 $  & $1.320 \pm 0.032 $  & $1.167 \pm 0.001 $  & $4.94 \pm 0.03 $  & $1.315 \pm 0.002 $  & $0.825 \pm 0.001 $  & $2.03 \pm 0.01 $  & $1.347 \pm 0.002 $ \\
 & 100 & $1.120 \pm 0.013 $  & $2.33 \pm 0.03 $  & $1.661 \pm 0.024 $  & $1.423 \pm 0.002 $  & $5.21 \pm 0.03 $  & $1.596 \pm 0.002 $  & $1.052 \pm 0.002 $  & $2.14 \pm 0.01 $  & $1.649 \pm 0.010 $ \\
 & 104 & $1.122 \pm 0.014 $  & $2.34 \pm 0.03 $  & $1.664 \pm 0.026 $  & $1.452 \pm 0.002 $  & $5.36 \pm 0.03 $  & $1.611 \pm 0.002 $  & $1.078 \pm 0.002 $  & $2.15 \pm 0.01 $  & $1.686 \pm 0.003 $ \\
 & 115 & $1.230 \pm 0.013 $  & $2.36 \pm 0.03 $  & $1.808 \pm 0.024 $  & $1.546 \pm 0.002 $  & $5.43 \pm 0.03 $  & $1.713 \pm 0.002 $  & $1.168 \pm 0.002 $  & $2.20 \pm 0.01 $  & $1.796 \pm 0.003 $ \\
 & 141 & $1.423 \pm 0.012 $  & $2.49 \pm 0.03 $  & $2.031 \pm 0.024 $  & $1.741 \pm 0.002 $  & $5.71 \pm 0.04 $  & $1.905 \pm 0.018 $  & $1.348 \pm 0.002 $  & $2.28 \pm 0.01 $  & $2.031 \pm 0.013 $ \\
\hline
3 & 50 & $0.445 \pm 0.013 $  & $2.07 \pm 0.04 $  & $0.714 \pm 0.024 $  & $0.634 \pm 0.001 $  & $4.60 \pm 0.02 $  & $0.726 \pm 0.001 $  & $0.438 \pm 0.001 $  & $1.81 \pm 0.01 $  & $0.786 \pm 0.001 $ \\
 & 52 & $0.452 \pm 0.006 $  & $2.03 \pm 0.01 $  & $0.738 \pm 0.010 $  & $0.646 \pm 0.001 $  & $4.64 \pm 0.02 $  & $0.738 \pm 0.001 $  & $0.450 \pm 0.001 $  & $1.82 \pm 0.01 $  & $0.801 \pm 0.005 $ \\
 & 58 & $0.507 \pm 0.006 $  & $2.08 \pm 0.02 $  & $0.813 \pm 0.011 $  & $0.687 \pm 0.001 $  & $4.67 \pm 0.02 $  & $0.784 \pm 0.001 $  & $0.485 \pm 0.001 $  & $1.85 \pm 0.01 $  & $0.856 \pm 0.005 $ \\
 & 71 & $0.558 \pm 0.002 $  & $2.12 \pm 0.01 $  & $0.881 \pm 0.003 $  & $0.772 \pm 0.001 $  & $4.73 \pm 0.02 $  & $0.880 \pm 0.001 $  & $0.556 \pm 0.001 $  & $1.89 \pm 0.01 $  & $0.960 \pm 0.002 $ \\
 & 100 & $0.723 \pm 0.009 $  & $2.10 \pm 0.03 $  & $1.151 \pm 0.018 $  & $0.939 \pm 0.001 $  & $4.86 \pm 0.02 $  & $1.068 \pm 0.010 $  & $0.711 \pm 0.001 $  & $1.99 \pm 0.01 $  & $1.178 \pm 0.007 $ \\
 & 104 & $0.718 \pm 0.009 $  & $2.10 \pm 0.02 $  & $1.143 \pm 0.017 $  & $0.959 \pm 0.001 $  & $4.93 \pm 0.02 $  & $1.080 \pm 0.001 $  & $0.727 \pm 0.001 $  & $1.99 \pm 0.01 $  & $1.206 \pm 0.007 $ \\
 & 115 & $0.793 \pm 0.009 $  & $2.13 \pm 0.03 $  & $1.250 \pm 0.018 $  & $1.020 \pm 0.001 $  & $4.95 \pm 0.02 $  & $1.149 \pm 0.001 $  & $0.788 \pm 0.001 $  & $2.03 \pm 0.01 $  & $1.285 \pm 0.002 $ \\
 & 141 & $0.910 \pm 0.008 $  & $2.20 \pm 0.02 $  & $1.399 \pm 0.017 $  & $1.144 \pm 0.001 $  & $5.08 \pm 0.03 $  & $1.286 \pm 0.002 $  & $0.909 \pm 0.001 $  & $2.08 \pm 0.01 $  & $1.449 \pm 0.009 $ \\
\hline
4 & 50 & $0.334 \pm 0.003 $  & $1.87 \pm 0.04 $  & $0.581 \pm 0.014 $  & $0.476 \pm 0.001 $  & $4.65 \pm 0.02 $  & $0.544 \pm 0.001 $  & $0.332 \pm 0.001 $  & $1.83 \pm 0.01 $  & $0.590 \pm 0.003 $ \\
 & 52 & $0.304 \pm 0.010 $  & $1.81 \pm 0.04 $  & $0.545 \pm 0.021 $  & $0.484 \pm 0.001 $  & $4.67 \pm 0.02 $  & $0.553 \pm 0.001 $  & $0.339 \pm 0.001 $  & $1.83 \pm 0.01 $  & $0.600 \pm 0.003 $ \\
 & 58 & $0.370 \pm 0.003 $  & $1.95 \pm 0.01 $  & $0.621 \pm 0.007 $  & $0.515 \pm 0.001 $  & $4.64 \pm 0.02 $  & $0.588 \pm 0.001 $  & $0.364 \pm 0.001 $  & $1.84 \pm 0.01 $  & $0.643 \pm 0.001 $ \\
 & 71 & $0.399 \pm 0.008 $  & $1.85 \pm 0.03 $  & $0.700 \pm 0.017 $  & $0.578 \pm 0.001 $  & $4.74 \pm 0.02 $  & $0.659 \pm 0.001 $  & $0.418 \pm 0.001 $  & $1.90 \pm 0.01 $  & $0.717 \pm 0.004 $ \\
 & 100 & $0.533 \pm 0.008 $  & $1.97 \pm 0.03 $  & $0.891 \pm 0.019 $  & $0.702 \pm 0.001 $  & $4.82 \pm 0.02 $  & $0.800 \pm 0.001 $  & $0.533 \pm 0.001 $  & $1.99 \pm 0.01 $  & $0.885 \pm 0.005 $ \\
 & 104 & $0.530 \pm 0.010 $  & $1.97 \pm 0.03 $  & $0.886 \pm 0.021 $  & $0.715 \pm 0.001 $  & $4.80 \pm 0.02 $  & $0.814 \pm 0.001 $  & $0.546 \pm 0.001 $  & $1.99 \pm 0.01 $  & $0.906 \pm 0.005 $ \\
 & 115 & $0.612 \pm 0.004 $  & $2.07 \pm 0.02 $  & $0.983 \pm 0.011 $  & $0.761 \pm 0.001 $  & $4.84 \pm 0.02 $  & $0.866 \pm 0.001 $  & $0.593 \pm 0.001 $  & $2.04 \pm 0.01 $  & $0.961 \pm 0.006 $ \\
 & 141 & $0.690 \pm 0.005 $  & $2.08 \pm 0.02 $  & $1.107 \pm 0.011 $  & $0.855 \pm 0.001 $  & $4.96 \pm 0.02 $  & $0.963 \pm 0.001 $  & $0.682 \pm 0.001 $  & $2.09 \pm 0.01 $  & $1.087 \pm 0.007 $ \\
\hline
5 & 50 & $0.241 \pm 0.009 $  & $1.75 \pm 0.04 $  & $0.446 \pm 0.020 $  & $0.380 \pm 0.001 $  & $4.68 \pm 0.02 $  & $0.434 \pm 0.001 $  & $0.267 \pm 0.001 $  & $1.84 \pm 0.01 $  & $0.472 \pm 0.001 $ \\
 & 52 & $0.254 \pm 0.019 $  & $1.77 \pm 0.08 $  & $0.466 \pm 0.041 $  & $0.388 \pm 0.001 $  & $4.66 \pm 0.02 $  & $0.443 \pm 0.001 $  & $0.274 \pm 0.001 $  & $1.85 \pm 0.01 $  & $0.481 \pm 0.001 $ \\
 & 58 & $0.291 \pm 0.004 $  & $1.81 \pm 0.01 $  & $0.521 \pm 0.008 $  & $0.412 \pm 0.001 $  & $4.70 \pm 0.02 $  & $0.470 \pm 0.001 $  & $0.364 \pm 0.001 $  & $1.84 \pm 0.01 $  & $0.643 \pm 0.001 $ \\
 & 71 & $0.350 \pm 0.003 $  & $1.89 \pm 0.02 $  & $0.604 \pm 0.008 $  & $0.462 \pm 0.001 $  & $4.73 \pm 0.02 $  & $0.527 \pm 0.001 $  & $0.336 \pm 0.001 $  & $1.90 \pm 0.01 $  & $0.577 \pm 0.001 $ \\
 & 100 & $0.432 \pm 0.007 $  & $1.87 \pm 0.03 $  & $0.752 \pm 0.016 $  & $0.562 \pm 0.001 $  & $4.84 \pm 0.02 $  & $0.639 \pm 0.001 $  & $0.429 \pm 0.001 $  & $2.00 \pm 0.01 $  & $0.706 \pm 0.001 $ \\
 & 104 & $0.421 \pm 0.005 $  & $1.85 \pm 0.01 $  & $0.739 \pm 0.011 $  & $0.572 \pm 0.001 $  & $4.85 \pm 0.02 $  & $0.651 \pm 0.001 $  & $0.439 \pm 0.001 $  & $2.00 \pm 0.01 $  & $0.723 \pm 0.004 $ \\
 & 115 & $0.455 \pm 0.007 $  & $1.84 \pm 0.02 $  & $0.804 \pm 0.016 $  & $0.608 \pm 0.001 $  & $4.82 \pm 0.02 $  & $0.692 \pm 0.001 $  & $0.474 \pm 0.001 $  & $2.03 \pm 0.01 $  & $0.773 \pm 0.005 $ \\
 & 141 & $0.538 \pm 0.007 $  & $1.95 \pm 0.02 $  & $0.907 \pm 0.014 $  & $0.683 \pm 0.001 $  & $4.95 \pm 0.02 $  & $0.769 \pm 0.001 $  & $0.547 \pm 0.001 $  & $2.09 \pm 0.01 $  & $0.872 \pm 0.001 $ \\
\hline
6 & 50 & $0.228 \pm 0.013 $  & $1.82 \pm 0.06 $  & $0.406 \pm 0.028 $  & $0.318 \pm 0.001 $  & $4.72 \pm 0.02 $  & $0.363 \pm 0.001 $  & $0.226 \pm 0.001 $  & $1.88 \pm 0.01 $  & $0.392 \pm 0.001 $ \\
 & 52 & $0.239 \pm 0.003 $  & $1.84 \pm 0.03 $  & $0.423 \pm 0.008 $  & $0.324 \pm 0.001 $  & $4.74 \pm 0.02 $  & $0.369 \pm 0.001 $  & $0.230 \pm 0.001 $  & $1.86 \pm 0.01 $  & $0.400 \pm 0.002 $ \\
 & 58 & $0.234 \pm 0.002 $  & $1.75 \pm 0.01 $  & $0.435 \pm 0.005 $  & $0.344 \pm 0.001 $  & $4.76 \pm 0.02 $  & $0.392 \pm 0.001 $  & $0.246 \pm 0.001 $  & $1.88 \pm 0.01 $  & $0.426 \pm 0.002 $ \\
 & 71 & $0.297 \pm 0.009 $  & $1.85 \pm 0.05 $  & $0.521 \pm 0.020 $  & $0.386 \pm 0.001 $  & $4.79 \pm 0.02 $  & $0.440 \pm 0.001 $  & $0.282 \pm 0.001 $  & $1.92 \pm 0.01 $  & $0.479 \pm 0.003 $ \\
 & 100 & $0.351 \pm 0.020 $  & $1.78 \pm 0.07 $  & $0.639 \pm 0.045 $  & $0.468 \pm 0.001 $  & $4.84 \pm 0.02 $  & $0.532 \pm 0.001 $  & $0.357 \pm 0.001 $  & $2.00 \pm 0.01 $  & $0.589 \pm 0.004 $ \\
 & 104 & $0.371 \pm 0.008 $  & $1.88 \pm 0.03 $  & $0.645 \pm 0.016 $  & $0.477 \pm 0.001 $  & $4.86 \pm 0.02 $  & $0.543 \pm 0.005 $  & $0.366 \pm 0.001 $  & $2.01 \pm 0.01 $  & $0.603 \pm 0.001 $ \\
 & 115 & $0.398 \pm 0.010 $  & $1.87 \pm 0.03 $  & $0.693 \pm 0.021 $  & $0.507 \pm 0.001 $  & $4.88 \pm 0.02 $  & $0.572 \pm 0.005 $  & $0.395 \pm 0.001 $  & $2.03 \pm 0.01 $  & $0.645 \pm 0.004 $ \\
 & 141 & $0.457 \pm 0.011 $  & $1.84 \pm 0.04 $  & $0.808 \pm 0.027 $  & $0.567 \pm 0.001 $  & $4.89 \pm 0.02 $  & $0.640 \pm 0.006 $  & $0.455 \pm 0.001 $  & $2.09 \pm 0.01 $  & $0.725 \pm 0.004 $ \\
\hline
\end{tabular}

		\caption{Student's $t$ distribution parameters $\nu$ and $\sigma$ fitted to the scattering angle distribution of $p =$~1-6~GeV/$c$ electrons on a $d =$~50-141~$\mu$m thin silicon target. For comparison, simulation results using the default Urb\'an model in Geant4 10.0 and our model are shown. In addition, the corresponding $RMS_{98}$ values are given.}
	\label{tab:MeasuredAndSimulatedValues}
\end{table*}

\bibliographystyle{unsrt_collab_comma}
\bibliography{mu3e}

\end{document}